# Cyclostationary Spectrum Sensing in Cognitive Radios Using FRESH Filters

Hemant Saggar[+] and D.K. Mehra[*]

Department of Electronics & Computer Engineering
Indian Institute of Technology Roorkee, Uttarakhand-247667
[+]hemantsaggar.iitr@gmail.com (corresp. author), [*]dkmecfec@iitr.ernet.in (Professor)

*Abstract: This paper deals with spectrum sensing in Cognitive Radios to enable unlicensed secondary users to opportunistically access a licensed band. The ability to detect the presence of a primary user at a low signal to noise ratio (SNR) is a challenging prerequisite to spectrum sensing and earlier proposed techniques like energy detection and cyclostationary detection have only been partially successful. This paper proposes the use of FRESH (FREquency SHift) filters [1] to enable spectrum sensing at low SNR by optimally estimating a cyclostationary signal using its spectral coherence properties. We establish the mean square error convergence of the adaptive FRESH filter through simulation. Subsequently, we formulate a cyclostationarity based binary hypothesis test on the filtered signal and observe the resultant detection performance. Simulation results show that the proposed approach performs better than energy detection and cyclostationary detection techniques for spectrum sensing.*

*Index Terms: Cognitive Radio, Cyclostationary Sensing, Energy Detection, FRESH filter, Spectrum Sensing.*

## I. INTRODUCTION

With a drastic increase in the number of wireless devices and services in the market accompanied by the demand for higher data rates, radio spectrum has become a precious and scarce resource. Notably, studies [2] have suggested that the licensed spectrum is used highly disproportionately, causing some bands to be overburdened while leaving others highly underutilized. Such unutilized frequency bands may be efficiently utilized by unlicensed users to transmit their information without disrupting the licensed user. Such an opportunity is called a *spectrum hole* and a device that can detect these holes and adapt its transmission parameters (frequency, modulation etc.) according to the changing RF environment is called a *Cognitive Radio* [3].

The task of detecting a spectrum hole across time and frequency, or equivalently the presence of a primary signal, known as *Spectrum Sensing* is not easy, given that such signals may be severely attenuated and heavily embedded in noise and co-channel interference. Different approaches like periodogram, energy detection, cyclostationary sensing [4],[5] have been proposed for this. The periodogram approach may not work reliably at low SNR due to energy smearing by windowing phenomenon. On the other hand energy detection is badly affected by uncertainties in noise variance [6]. Instead, by exploiting the second order periodicity inherent in most modulated signals (see section II), cyclostationary detection can be used for reliable spectrum sensing. Cyclostationary sensing has been suggested in the past [7] and combined with other approaches like neural networks [8] and is a more reliable method of spectrum sensing at low SINR.

FRESH or Frequency Shift filters were proposed by Gardener [1] as the optimum time varying filter structures for estimation of cyclostationary signals. They consist of a bank of filters, each preceded by a frequency shifter that is tuned to a specific periodicity in the incoming cyclostationary signal. Hence FRESH filtering is also called polyperiodic filtering. Adaptive versions of these filters both blind and otherwise, using LMS and RLS algorithms were proposed in [9]. Recently the authors in [10] have studied the effect of cycle frequency errors on MSE of FRESH filters.

In this paper, we approach FRESH filters from a perspective of spectrum sensing and propose to apply them to enable sensing in the low SNR regime. We build a blind adaptive LMS-FRESH filter to detect a BPSK signal with known cyclic frequencies buried in AWGN and show that the filter attains convergence in mean square error sense. In order to complete the task of spectrum sensing, a cognitive user must make a decision based on the output of the FRESH filter as to whether the primary signal is present or not. As stated earlier, using energy detection yields a highly unreliable test at low SNR due to noise uncertainty problem which may also fail to limit the false alarm rate. To counter this problem, we propose to measure the Conjugate Cyclic Autocorrelation function of the filter output and formulate a binary hypothesis test with a constant false alarm rate. Simulation results show that our approach of spectrum sensing using FRESH filter and cyclostationarity based hypothesis test, gives better detection performance as compared to energy detection and cyclostationary detection without FRESH filters.

Section II discusses energy detection and cyclostationary sensing and section III explains the theory behind FRESH filters. Section IV presents the setup used and results of the simulation and finally the conclusions are presented in section V.

## II. SPECTRUM SENSING

Spectrum Sensing is the task of obtaining awareness about the spectrum usage and existence of primary users in a given frequency band, though it can also be extended to other dimensions like space, code and angle [5]. This is a key functionality of any Cognitive radio that ensures minimal interference towards the primary occupant of the band and maximizes the transmission capacity of secondary user. But, due to attenuation, shadowing and co channel interference [5], a cognitive radio must operate at very low SNR, necessitating high resolution ADC/DAC's and faster DSP's. Additionally, stringent limits on resultant primary user interference and channel sensing duration [11] defined in standards such as IEEE 802.22 have put the bar very high for potential spectrum sensing algorithms. Below we discuss two major algorithms that have been proposed for spectrum sensing [4],[5].

### A. Energy Detection

Radiometry or energy detection is the simplest and fastest method of spectrum sensing having low computation complexity. An energy detector can be modeled as follows. Consider the samples of a complex signal $x(n)$ of zero mean and variance $\sigma_x^2$ being transmitted through an AWGN channel with noise variance $\sigma_w^2$. Then the received signal $r(n)$ can be written as

$$\begin{aligned} H_0: \quad & r(n) = w(n) \\ H_1: \quad & r(n) = x(n) + w(n) \end{aligned} \quad (1)$$

The test statistic used, is the energy of the received signal measured as

$$\mathbb{E} = \sum_{i=1}^{N}[r(n)] \times [r(n)]^* \quad (2)$$

where $[.]^*$ denotes complex conjugation and $N$ is the number of received samples. If we fix the false alarm probability ($P_f$) and based on the corresponding decision threshold λ, calculate the detection probability ($P_d$), then by using central limit theorem we obtain the following relations [4],

$$P_f = Q\left(\frac{\lambda - N\sigma_w^2}{\sqrt{2N\sigma_w^4}}\right) \quad (3)$$

$$P_d = Q\left(\frac{\lambda - N(\sigma_w^2 + \sigma_x^2)}{\sqrt{2N(\sigma_w^2 + \sigma_x^2)^2}}\right) \quad (4)$$

Unfortunately, at low SNR these formulae do not hold valid due to a phenomenon called SNR wall [6]. Due to uncertainty in measured noise variance, the energy detector cannot limit the false alarm rate and fails to detect primary signal below a specific SNR irrespective of the number of samples $N$ used. This uncertainty in noise variance can be modeled as,

$$\begin{aligned}\sigma_{w_1}^2 &= \sigma_w^2 * 10^{-a/10} \\ \sigma_{w_2}^2 &= \sigma_w^2 * 10^{a/10}\end{aligned} \quad (5)$$

where $a$ dB is the uncertainty in noise variance. The worst case detection performance can be calculated by using $\sigma_{w_1}^2$ to set $P_f$ and $\sigma_{w_2}^2$ to find the resultant $P_d$ (see figures 3-5).

### B. Cyclostationary Sensing

Cyclostationary signals exhibit second order periodicity, hence we can define their Cyclic Autocorrelation Function (CAF) [1]

$$R_{xx^{[*]}}^\alpha(k) = \sum_{n=1}^{N} R_{xx^{[*]}}(n, n+k)e^{-j2\pi\alpha nf_s} \quad (6)$$

where $x(n)$ is the input signal, $\alpha$ is the cyclic frequency, $k$ is the lag value and $[*]$ denotes the optional conjugation to obtain the conjugate CAF, $R_{xx^*}^\alpha(k)$. We again assert that since $x(n)$ is complex its autocorrelation is found as

$$R_{xx}(n, n+k) = E\{x(n)x^*(n+k)\} \quad (7)$$

A signal $x(n)$ may possess multiple cyclic frequencies ($\alpha$) at which the CAF exhibits a peak. Consider the signal

$$x(t) = a(t)\cos(\omega_c t) + ib(t)\sin(\omega_c t) \quad (8)$$

$$a(t) = \sum_{r=-\infty}^{\infty} c_r p(t - rT_0) \quad (9)$$

where $\omega_c = 2\pi f_c$, $p(t)$ is the baseband rectangular pulse, $T_0$ the baud rate and $c_r = \pm 1$. If $a(t)$ and $b(t)$ are proportional then $x(t)$ becomes a BPSK signal. The cyclic frequencies of the BPSK signal with k=±1, ±2,… are [1]

$$\begin{aligned}R_{xx}^\alpha(k) &\neq 0 \text{ for } \alpha = \pm k/T_0 \\ &= 0 \text{ for } \alpha \neq \pm k/T_0 \\ R_{xx^*}^\alpha(k) &\neq 0 \text{ for } \alpha = \pm 2f_c \pm \frac{k}{T_0} \\ &= 0 \text{ for } \alpha \neq \pm 2f_c \pm \frac{k}{T_0}\end{aligned} \quad (10)$$

### III. FRESH FILTERS

Frequency Shift or FRESH filters are time varying filters which exploit the spectral coherence in cyclostationary signals to optimally estimate them. As the name suggests, a FRESH filter consists of many branches, each having a frequency shifter followed by a time invariant filter where each shift is equal to the cyclic frequency of the input signal. That is why they are also called LAPTV (Linear Almost Periodic Time Varying) filters. Depending on the input signal, both conjugate and non-conjugate cyclic frequencies are used.

The output $y(n)$ of a LCL-FRESH (Linear Conjugate Linear) filter with input $x(n)$ can be written as

$$y(n) = \sum_{p=1}^{M} \boldsymbol{a}_p(n) \otimes \boldsymbol{x}_{\alpha_p}(n) \\ + \sum_{q=1}^{N} \boldsymbol{b}_q(n) \otimes \boldsymbol{x}^*_{-\beta_q}(n) \quad (11)$$

where $\{\alpha_p\}$ and $\{\beta_q\}$ are cyclic frequencies of the input signal, $\boldsymbol{a}_p(n)$ and $\boldsymbol{b}_q(n)$ the weight vectors of the FIR filters of the $p^{th}$ ($p=1$ to $M$) linear and $q^{th}$ ($q=1$ to $N$) conjugate linear branch respectively.

$$\boldsymbol{x}_{\alpha_p}(n) \\ \triangleq \boldsymbol{x}(n).\{\exp(i2\pi\alpha_p nT_s)\}_{n=1:M_p} \quad (12)$$

$$\boldsymbol{x}^*_{-\beta_q}(n) \\ \triangleq \boldsymbol{x}^*(n).\{\exp(i2\pi\beta_q nT_s)\}_{n=1:N_q} \quad (13)$$

Here $(.)$ denotes element-wise product, $M_p$ and $N_q$ are the lengths of $p^{th}$ and $q^{th}$ FIR filters respectively and $T_s$ is the sample duration. Blind adaptive FRESH filters have been designed by authors in [9]. Consider the concatenated weight vector $\boldsymbol{w(n)}$ equal to $\{\boldsymbol{a}_1(n) \ldots \boldsymbol{a}_M(n), \boldsymbol{b}_1(n) \ldots \boldsymbol{b}_N(n)\}$. Then the weight update equations can be written as follows,

$$\epsilon(n) = y(n) - x(n) \\ \boldsymbol{w}(n+1) = \boldsymbol{w}(n) + \mu\epsilon^*(n)\boldsymbol{x}(n) \quad (14)$$

## IV. SIMULATION RESULTS

The transmitted signal is assumed to be a complex BPSK signal of the form $x(t) = s(t)\exp(j2\pi f_c t)$ being passed through an AWGN channel with a unit noise variance ($\sigma_w^2$). $s(t)$ is a random NRZ binary data sequence with full duty cycle square pulses of duration $T_0$. The carrier frequency, $f_c$, and baud rate, $\frac{1}{T_0}$, are chosen as 30720 Hz and 3200 Hz respectively. At the receiver, the signal $r(n)$ is the sampled value of $r(t) = x(t) + w(t)$, sampled at a sufficiently high rate to avoid aliasing at cyclic frequencies close to $2f_c$. This signal is then passed through a FRESH filter as shown in figure 1.

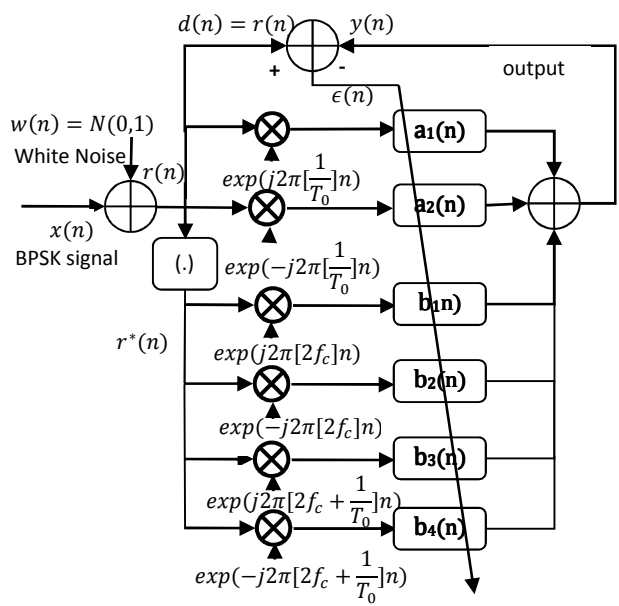

Fig. 1. LMS blind adaptive FRESH filter with six branches for estimating a BPSK signal.

Three cyclic frequencies of the BPSK signal $x(n)$ are utilized in the FRESH filter [1], $\alpha_1 = \frac{1}{T_0}, \alpha_2 = 2f_c$ and $\alpha_3 = 2f_c + \frac{1}{T_0}$. The six branches of the filter correspond to frequency shifts of $\alpha = \{\pm\alpha_1, \pm\alpha_2, \pm\alpha_3\}$ followed by filtering by a 64 tap FIR filter whose weights are adapted through Least Mean Square (LMS) algorithm. The sampling frequency $f_s$ is chosen to be 32 times the baud rate. The received signal is used as the training signal. The estimation error is $\epsilon(n) = r(n) - y(n)$. Through simulation, a suitable convergence factor (μ) is chosen that minimizes the mean square error (MSE) $\mathbb{E}\{\|\epsilon(n)\|^2$ as shown in figure 2.

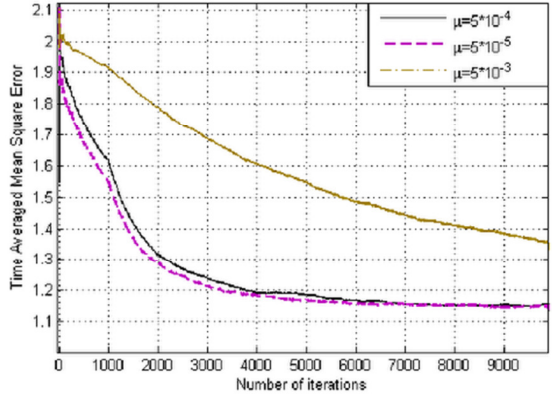

Fig. 2. Variation of time averaged MSE with number of iterations for different μ. SNR=0dB.

A convergence factor $\mu = 5 * 10^{-5}$ is found to give the minimum MSE over a wide range of SNR. Now, we construct a test statistic T equal to the sum of Conjugate Cyclic Autocorrelation of $y(n)$ at $\alpha = 2f_c, 2f_c + \frac{1}{T_0}, 2f_c - \frac{1}{T_0}$ and formulate a binary hypothesis test as

$$\mathbb{T} = R_{yy^*}^{2f_c}(k) + R_{yy^*}^{2f_c+\frac{1}{T_0}}(k) + R_{yy^*}^{2f_c-\frac{1}{T_0}}(k) \quad (15)$$
$$P_f = \Pr\{T > \lambda | H_0\}$$
$$P_d = \Pr\{T > \lambda | H_1\}$$

where $\lambda$ is the decision threshold. We consider only the stronger cyclostationarity of BPSK at $2f_c$. It is important to note that under the null hypothesis, stationary noise $w(n)$ passing through the FRESH filter is converted to cyclostationary noise [12], hence $\Pr\{\mathbb{T}|H_0\} \neq 0$.

To obtain the average detection performance of this scheme, we simulate 1000 independent runs from SNR 0 dB to -20 dB with $P_f = 0.01$ and calculate the average detection rate. Figures 3-5 compare the observed detection performance of the proposed scheme with that of energy detection with known and unknown noise variance (± 1dB uncertainty) and cyclostationary detection without using FRESH filter. For the latter case, hypothesis test of (15) is applied directly to the received signal. It is observed that FRESH filter aided cyclostationary sensing gives significantly better detection performance than all other techniques in the low SNR regime. As the number of samples available increases, the detection performance improves concurrently.

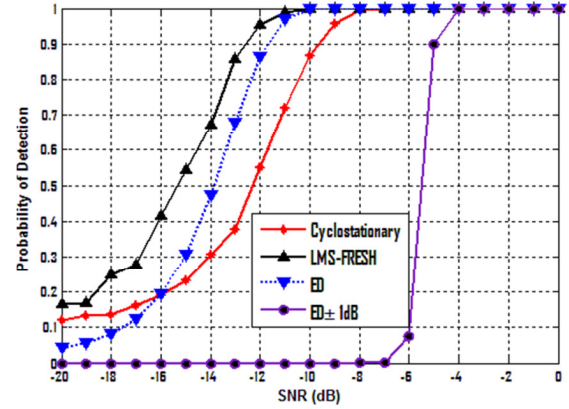

Fig. 3. Probability of detection versus SNR(dB) for 800 samples.

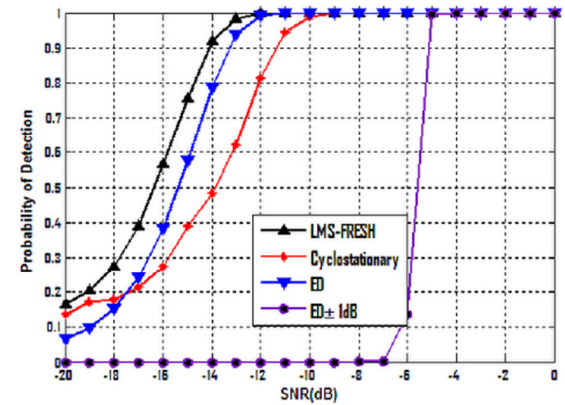

Fig. 4. Probability of detection versus SNR(dB) for 1600 samples.

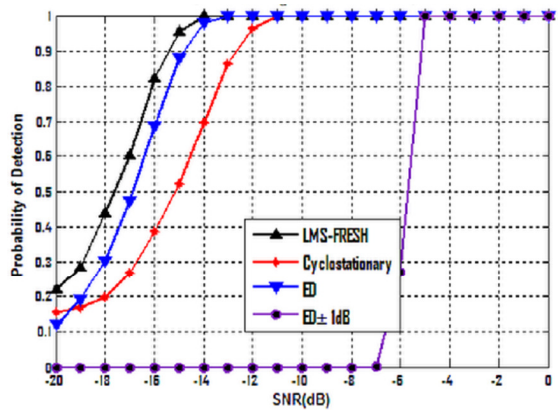

Fig. 5. Probability of detection versus SNR(dB) for 3200 samples

## V. CONCLUSION

In this paper we propose the application of FRESH filters for spectrum sensing in Cognitive Radios. We use the FRESH filter structure given in section IV to detect a cyclostationary signal and propose a cyclostationarity based hypothesis test on the filter output to enable spectrum sensing. Simulation results show that the proposed approach gives better detection probability in the low SNR regime as compared to energy detection. We also show that the use of FRESH filters can improve the detection probability in cyclostationary spectrum sensing.